\documentstyle[preprint,aps,prl,epsf]{revtex}
\begin{document}

\draft

\title{Ground-State Magnetization in Disordered Systems : \\
Exchange vs. Off-Diagonal Interaction Fluctuations}
\author{Philippe Jacquod and A. Douglas Stone}
\address{Applied Physics Department, P.O. Box 208284 \\
Yale University, New Haven, CT 06520-8284}

\maketitle
\begin{abstract}
\noindent In a Hartree-Fock picture, itinerant ferromagnetism results from
a competition between kinetic and exchange energy, with the magnetized
state being favored at large interaction strength.
In a recent paper \cite{spings} we showed that in contrast to this average
effect,
fluctuations of off-diagonal interaction matrix elements systematically
reduce the ground-state magnetization. When the interaction dominates,
the occurence of a non-zero ground-state magnetization depends on the
ratio $\lambda$ between average exchange energy and the fluctuation
amplitude of the
off-diagonal matrix elements, and a nonzero critical
value $\lambda_c$ is necessary to magnetize the ground-state.
We extend these results by numerical studies
of $\lambda$ for standard tight-binding models which
indicate a regime of intermediate
disorder where off-diagonal fluctuations should play an important
role for ground-state magnetization. We also emphasize the presence of
strong correlations between minimal eigenvalues of different
magnetization blocks which further reduce the probability
of a nonzero ground-state spin.
\end{abstract}
\bigskip
\noindent Itinerant ferromagnetism results from the interplay between the
Pauli principle and the electronic interactions.
The exchange energy can be minimized when the fermionic
antisymmetry is supported by the spatial wavefunction and when the
resulting energy gain exceeds the corresponding loss in kinetic
energy, this implies spin alignment and a nonzero ground-state magnetization.
Because of the locality of the Pauli principle, ferromagnetism is often
studied in the framework of the Hubbard model for which a simple
Hartree-Fock picture predicts ferromagnetism to occur when the {\it Stoner
criterion}\cite{stoner} is satisfied $U_c \rho(\epsilon_F)=1$, where
$\rho(\epsilon_F)$ is the one-particle density of states at the Fermi energy.
Within perturbation theory, the effective interaction strength is enhanced
by the presence of disorder \cite{altshuler} and this had led to recent
predictions of
the occurence of a ferromagnetic instability already below
the Stoner threshold \cite{andreev}. However as we recently pointed
out, there is a competing effect of the interaction manifested through the
fluctuations of off-diagonal interaction
matrix elements which suppresses magnetization\cite{spings}. Indeed as is
known from studies of models in nuclear physics,
strong off-diagonal fluctuations can determine
the variance of the Many-Body Density of States (MBDOS) \cite{brody}.
When one includes the spin degree of freedom within such models, the
Hamiltonian splits into
blocks of different total magnetization $\sigma_z$ and one finds that
these fluctuations result in a broader MBDOS for the block with smallest
magnetization, increasing thereby the probability of finding a non
magnetized ground-state; therefore the average exchange effect must compete
not just with the kinetic energy but with off-diagonal fluctuations.

\noindent To shortly summarize the results
presented in \cite{spings}, we start from
a Hamiltonian for $n$ spin-$1/2$ particles distributed over $m/2$ one-particle
orbitals $\epsilon_{\alpha} \in [-m/2;m/2]$ (the mean level spacing is then
fixed as $\Delta \equiv 1$ with spin degeneracy)

\begin{eqnarray}
H & = & H_0 + H_U + H_S= \sum \epsilon_{\alpha}
c^{\dagger}_{\alpha,s} c_{\alpha,s} + \sum
  U_{\alpha,\beta}^{\gamma,\delta}
c^{\dagger}_{\alpha,s} c^{\dagger}_{\beta,s'} c_{\gamma,t} c_{\delta,t'}
- \lambda U \sum \vec{s}_{\alpha} \vec{s}_{\beta}
\end{eqnarray}

\noindent This Hamiltonian is a generic model of interacting
fermions expressed in the basis of Slater determinants constructed from
the eigenstates $\psi_{\alpha}$ of the corresponding free fermions model $H_0$.
The interaction matrix elements are given by
$ U_{\alpha,\beta}^{\gamma,\delta} = \int d\vec{r} d\vec{r'}
U(\vec{r}-\vec{r'}) \psi_{\alpha}^*(\vec{r})  \psi_{\beta}^*(\vec{r'})
  \psi_{\gamma}(\vec{r})  \psi_{\delta}(\vec{r'})$, where
$U(\vec{r}-\vec{r'})$ is the interaction potential.
Disorder results in a random character of
the $\psi_{\alpha}$'s, leading to fluctuations
in $U_{\alpha,\beta}^{\gamma,\delta}$ around their average value.
Only diagonal matrix elements have a nonzero average
leading to mean-field charge-charge and spin-spin
diagonal interactions. We neglect in (1) the mean-field
charge-charge contribution since it has
no influence on the magnetization and assume gaussian-distributed
$ U_{\alpha,\beta}^{\gamma,\delta}$ :
$ P(U_{\alpha,\beta}^{\gamma,\delta}) \propto
e^{-(U_{\alpha,\beta}^{\gamma,\delta})^2/2U^2}$.
$\vec{s}_{\alpha} \equiv \sum_{s,t}
c^{\dagger}_{\alpha,s} \vec{\sigma}_{s,t} c_{\alpha,t}$ are spin operators
and the parameter $\lambda > 0$
is given by $ \lambda \equiv \langle
U_{\alpha,\beta}^{\beta,\alpha} \rangle/\left(2 {\rm RMS}
\left[U_{\alpha,\beta}^{\gamma,\delta}\right]
\right)$, i.e. it is half the ratio
of the average exchange interaction to the root mean square value
of the off-diagonal matrix elements.
Finally, the interaction commutes with the total magnetization so
that the second sum on the right-hand side of (1)
is restricted to spin indices satisfying $s+s'=t+t'$. The
Hamiltonian acquires a block structure where blocks are labelled by their
total magnetization.
Each block's size is given in term of binomial coefficients as
$N(\sigma_z) = \left(_{n/2-\sigma_z}^{m/2}\right)
\left(_{n/2+\sigma_z}^{m/2}\right)$. Most importantly,
the two-body interaction we consider connects only
Slater determinants which differ by at most two
one-particle indices and each magnetization block's {\it connectivity}
(the number of nonzero matrix elements per
row) $K(\sigma_z)$ is a monotonously decreasing function of $\sigma_z$.

\noindent Similarly as in \cite{brody} we estimate the variance of each
magnetization block's MBDOS as

\begin{equation}\label{dosvar}
\frac{1}{N(\sigma_z)} \sum_{I,J} H_{I,J}^2 \delta(\sigma_I-\sigma_z)
\approx K(\sigma_z) U^2
\end{equation}

\noindent where $ H_{I,J} = \langle I| H |J \rangle $ and
$|I \rangle$ is a Slater determinant. Hence each block's bandwidth goes
as $\sqrt{K(\sigma_z)} U$. Moreover, earlier work tells us that in such
models as (1), the MBDOS is
approximately gaussian with vanishing corrections in the dilute
limit $1 \ll n \ll m$ \cite{brody}. We can therefore
conclude that {\it the full MBDOS is a sum of approximately gaussian
contributions from each
spin block with a variance proportional to the corresponding connectivity.
The latter is a monotonously decreasing function of $\sigma_z$,
hence the broadest MBDOS corresponds to the minimally magnetized block
and the ground state will be found in this block with increased probability.}
For approximately gaussian distributions one may assume that the tail of
each block's MBDOS scales with its variance
(up to a numerical factor $\beta$). Neglecting contributions from $H_0$,
the typical spin gap can be estimated (for $\lambda = 0$) as

\begin{equation}\label{spingap}
\Delta_s^{U} \approx \beta U[ \sqrt{K(|\sigma_{min}|)} -
\sqrt{K(|\sigma_{min}| + 1)}]
\end{equation}

\noindent Switching on the mean-field spin-spin interaction $H_S$
induces shifts of the block's MBDOS by energies of at most
$- \lambda U \sigma_z^2$
and the spin gap becomes on average

\begin{equation}\label{eq:gap}
\langle \Delta_s \rangle = \Delta_s^{U} - \bar{\lambda} U
\end{equation}

\noindent Results presented in \cite{spings} indicate only a very weak
$m$-dependence of $\Delta_s^{U}$. It has however
a strong even-odd dependence : $\Delta_s^U \approx 4U$ (8$U$)
for even (odd) number $n$ of electrons.
This effect is nevertheless compensated by a twice larger antiferromagnetic
shift between the two lowest magnetized blocks for odd $n$ ($\sigma_z=1/2$
and $3/2$) than for even $n$ ($\sigma_z=0$ and $1$). Hence we introduced
$\bar{\lambda}=(3-(-1)^n)\lambda/2$ in (\ref{eq:gap})
which results in a parity independent behavior of $\lambda$.
Numerically we found a critical value of $\lambda_c \approx 2$
to have a non-negligible probability of magnetizing the ground-state. A larger
value $\approx 4$ is then necessary to have $\langle \Delta_s \rangle = 0$
resulting in a 50$\%$ probability of nonvanishing magnetization.

\noindent These points have been studied in details in \cite{spings}.
Here we first address
the fact that equation (\ref{eq:gap}) gives the average value of the spin gap,
based on the assumption that the lowest levels $E_0$ and $E_1$
in the two blocks with lowest magnetization (they define the spin
gap $\Delta_s \equiv E_1 - E_0$) have {\it uncorrelated distributions}.
To check this we study the following cross-correlation function

\begin{equation}
R(E_0,E_1) = \frac{\langle (E_0-\langle E_0 \rangle) (E_1-\langle E_1 \rangle)
\rangle}{\sqrt{\langle (E_0-\langle E_0 \rangle)^2
\rangle \langle (E_1-\langle E_1 \rangle)^2 \rangle}}
\end{equation}

\noindent It is easily seen that if $E_0$ and $E_1$ fluctuate independently
around their disorder averages $\langle E_0 \rangle$ and $\langle E_1 \rangle$,
$R(E_0,E_1) \to 0$
whereas strong correlations between $E_0-\langle E_0 \rangle$ and
$E_1-\langle E_1 \rangle$ - i.e. if a large (small) value of one of
them corresponds
to a large (small) value of the other - implies $R(E_0,E_1) \approx 1$. Fig.1
shows the $U$-dependence of $R(E_0,E_1)$ for the Hamiltonian (1) and
$n$=4, 5 and 6 electrons on $m=10$ orbitals
(the function is independent on $\lambda$). The numerical data show
indeed strong correlations between extremal levels of the two lowest
magnetized blocks. Even though $R(E_0,E_1)$ is somehow reduced by $H_U$,
correlations still remain quite strong $R(E_0,E_1)>0.8$ up to very large
$U=10$. This is understandable if we remember that all blocks, having
$K(\sigma_z) N(\sigma_z)$ nonzero matrix elements, are constructed out of
the same set of only $(m(m-1)/2)^2$ different two-body interaction matrix
elements. Extremal eigenvalues in the distribution are then due to special
realizations of
the latter inside the blocks. These realizations are presumably
not very different in blocks with consecutive magnetization which
results in strong eigenvalues correlations.
The existence of these strong correlations is confirmed by the
gap distribution $P(\Delta_s^U)$
for $H_U$ plotted on Fig. 2. Indeed the
probability to find a zero or negative gap is vanishingly small
($\int^0_{-\infty}P(\Delta_s^U) d(\Delta_s^U)=0$ for $n=5$ and $<0.002$
for $n=4$ and 6), whereas
even at large average gap, it would not vanish in the case
of two independent distributions for $E_0$ and $E_1$.
Fig. 2 indicates an effective repulsion between extremal eigenvalues
of different random matrices, built however from the same (small) set
of matrix elements. This repulsion results in a further decrease of the
probability for a magnetized ground-state.
Note that the even-odd dependence of the average of these distributions is
removed
due to the horizontal scaling $\Delta_s^U/\langle \Delta_s^U \rangle$.

\noindent We now turn our attention to the microscopic computation of
the magnetization parameter $\lambda$ for realistic solid-state models.
Specifically, we concentrate on the two-dimensional Anderson lattice

\begin{eqnarray}
H & = & V \sum_{\langle i;j \rangle} c^\dagger_{i,s} c_{j,s}
+ \sum_i W_i c^\dagger_{i,s} c_{i,s}
\end{eqnarray}

\noindent where $\langle i;j \rangle$ restricts the sum to nearest neighbors,
$W_i \in [-W/2;W/2]$ and $W$ is the disorder strength.
Also we consider different interaction potentials of the form

\begin{equation}
U(\vec{r}-\vec{r'}) = U_0 \delta(\vec{r}-\vec{r'}) +
U_1/\left| \vec{r}-\vec{r'} \right|
\end{equation}

\noindent Fig. 3 shows the disorder dependence of $\lambda$,
for the Hubbard interaction
case $U_1=0$ and different linear system size $L$=10, 20, 50 and 80.
The data have been obtained
from averages over 30 wavefunctions in the middle of the band $E =0$
and for 10 ($L=80$) to 200 ($L=10$) disorder realizations.
Clearly, three
regimes are distinguishable. ($I$) At low disorder, the one-electron
dynamics undergoes a crossover from ballistic to diffusive regime
as the linear system size is increased beyond the elastic mean-free path
$l_e \sim 50 (V/W)^2$. In the ballistic regime $l_e \gg L$,
wavefunctions are plane-waves. In this case, a Hubbard interaction
gives $\lambda \sim L^2$, since the
${\rm RMS} \left[U_{\alpha,\beta}^{\gamma,\delta}\right] \sim L^{-4}$ and
$\langle
U_{\alpha,\beta}^{\beta,\alpha} \rangle \sim L^{-2}$, whereas
once the diffusive regime
is reached, one expects $\lambda \sim \Delta/(\Delta/g) \sim g$ ($g$ is
the conductance) \cite{altshuler,blanter}.
($II$) In the regime of intermediate disorder, both
off-diagonal fluctuations and exchange are increased by disorder, and
apparently they
compensate each other, resulting in a $L$-independent $\lambda \approx 4$,
i.e. close to the critical value to have a
vanishing average gap (\ref{eq:gap}).
Therefore we can expect that
in this regime, and even for very strong
interaction, the probability for finite ground-state magnetization
does not exceed 50$\%$.
Note that for $W/V=5$, the localization length
is slightly below the largest system size considered and one can expect
finite-size effects to play only a minor role. ($III$) In the regime of strong
disorder, one-particle wavefunctions are strongly localized on fewer
and fewer sites, so that many of the interaction matrix elements
are much smaller than $U$, the off-diagonal fluctuations are
sharply reduced and again exchange dominates. Note that eventually, the
latter disappears also,
but at a lower rate than the fluctuations.

We finally evaluate the influence of the interaction range by considering
$U_1 \ne 0$. The average exchange interaction term is explicitly given by

\begin{equation}
\langle \int d\vec{r} d\vec{r'}
U(\vec{r}-\vec{r'}) \psi_{\alpha}^*(\vec{r})  \psi_{\beta}^*(\vec{r'})
  \psi_{\beta}(\vec{r})  \psi_{\alpha}(\vec{r'}) \rangle_{\alpha,\beta}
\end{equation}
$\langle ... \rangle_{\alpha,\beta}$ is an average taken over one-particle
wavefunctions close to the Fermi level. Due to their orthogonality,
taking this average over the full set
of wavefunctions gives a $\delta$-function so that only the on-site term
contributes. This last procedure is exact if the one-body dynamics
is described by Random Matrix Theory (RMT) for which the structure of the
eigenstates is homogeneous all through the spectrum.
In this case - which is relevant for irregular quantum dots
with chaotic scattering at the boundary \cite{jala} - only the on-site
part of the interaction contributes to the exchange, whereas it
is expectable that increasing the strength $U_1$ of the long-range terms
leads to stronger off-diagonal fluctuations, and a reduction of $\lambda$.
For the Anderson model however,
one-particle wave-functions are different from RMT so that
the average over wavefunctions close to the Fermi level leads only to a
more or less sharply peaked function of $(\vec{r}-\vec{r'})$ : there are also
contributions to the exchange from
the long-range terms, but still we expect that the average damps
them with respect to their contribution to off-diagonal fluctuations
(This damping of course depends on the disorder strength.)
so that $\lambda$ should decrease with increasing interaction range. The
validity of this reasoning is illustrated on Fig. 4
where we plot the evolution of
$\lambda$ for different disorders as the long-range part of the interaction
becomes more and more important. Clearly, $\lambda$ is increased by the
screening of the electron-electron interaction, and therefore the Hubbard
results presented on Fig. 3 give an upper boundary for $\lambda$.
One thus expects the demagnetizing effect described here and
in \cite{spings} to be more efficient at low filling when the screening length
exceeds the elastic mean free path.

Numerical computations were performed at the Swiss Center
for Scientific Computing. Work supported by the NSF grant PHY9612200 and
the Swiss National Science Foundation. It is a pleasure to acknowledge
interesting discussions with S. {\AA}berg, I. Aleiner, Y. Alhassid,
F. Izrailev and S. Tomsovic.

\nopagebreak

\newpage

\begin{figure}
\epsfxsize=6.2in
\epsfysize=5.4in
\epsffile{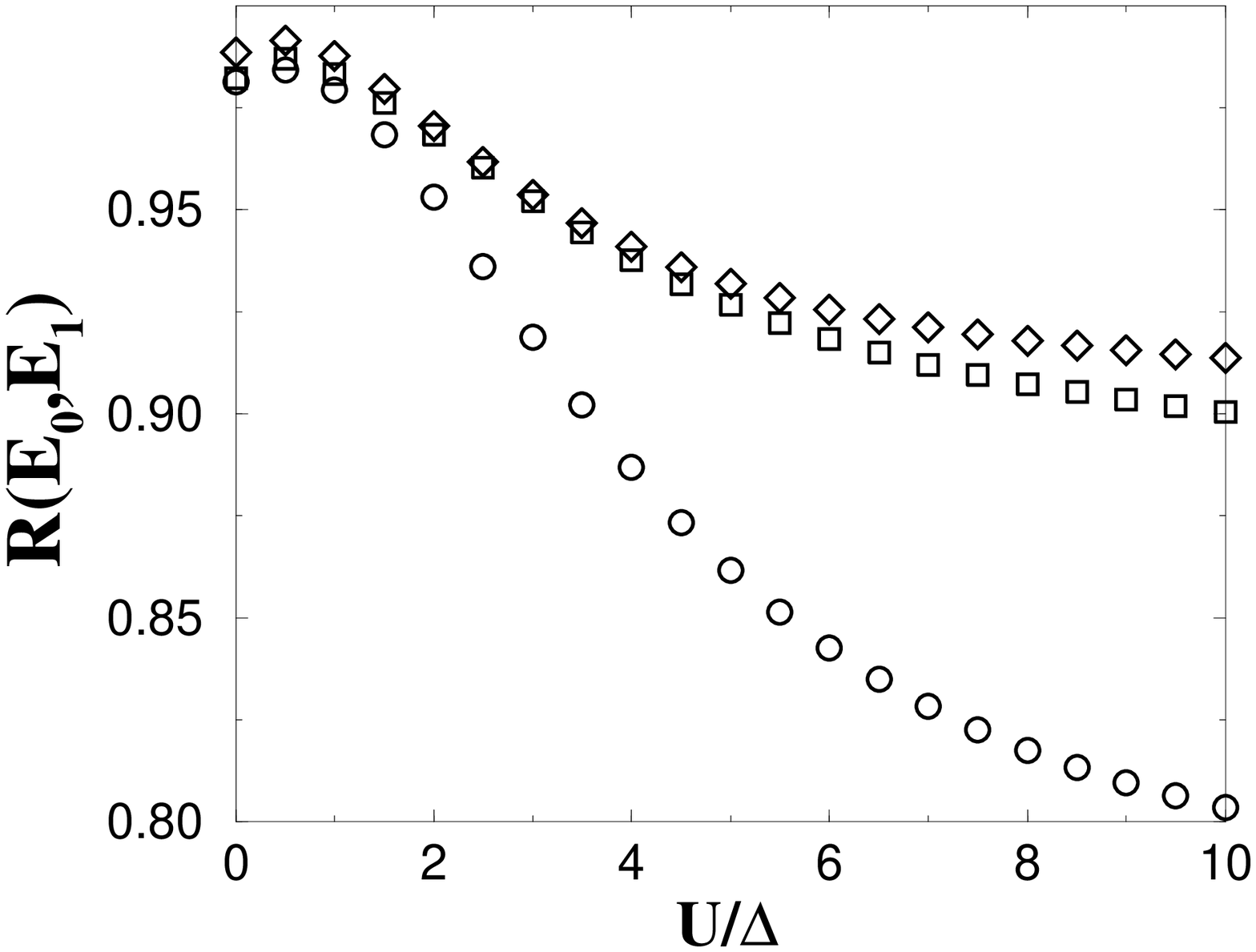}
\caption{Cross-correlation function (5) between the lowest energy levels
in the two blocks with lowest magnetization for 1000 realisations
of the Hamiltonian (1) with
$n$=4 (circles), 5 (squares) and 6 (diamonds) particles on $m=10$ orbitals.}
\label{fig:correlation}
\end{figure}

\begin{figure}
\epsfxsize=6.2in
\epsfysize=5.4in
\epsffile{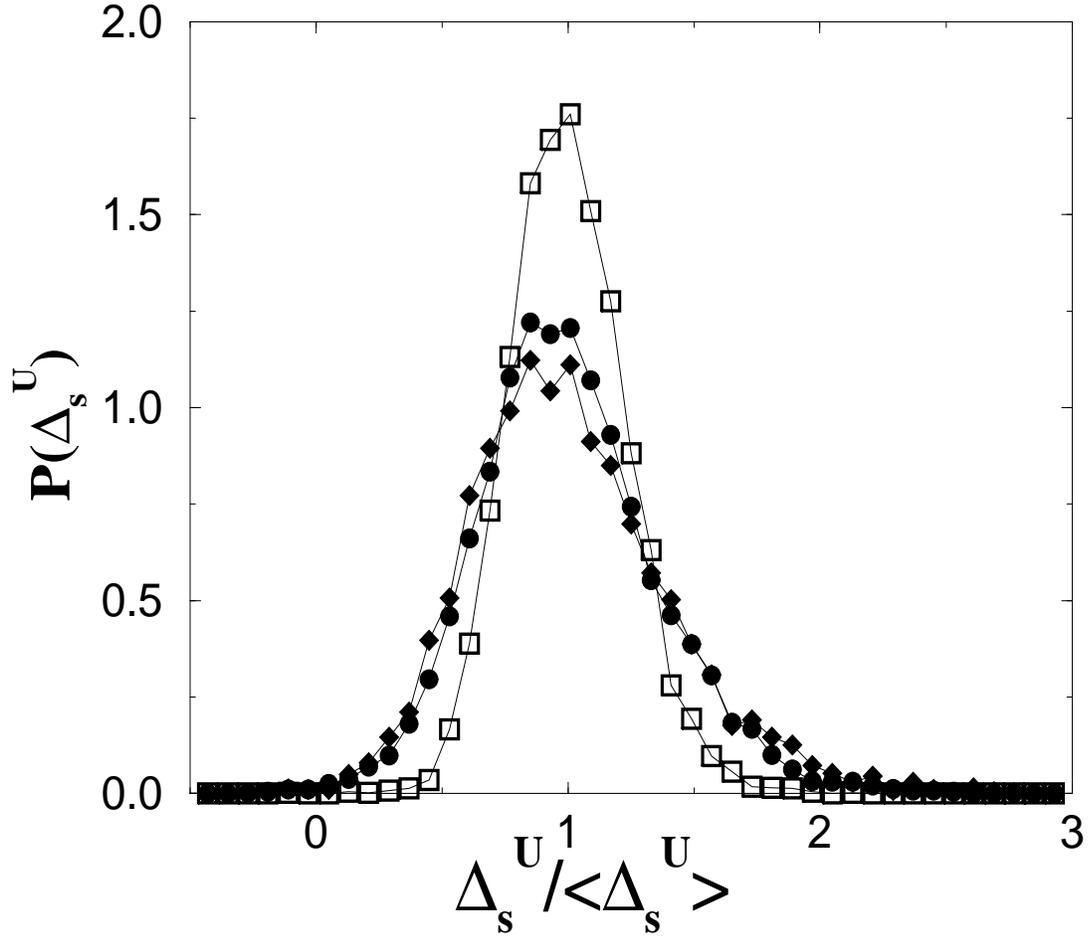}
\caption{Distribution of the spin gap $\Delta_s^U$ for $n$=4 (full circles),
5 (empty squares) and 6 (diamonds)
particles on $m=10$ orbitals. The distributions have been constructed
from 20000 realisations of the Hamiltonian $H_U$.}
\label{fig:repulsion}
\end{figure}

\begin{figure}
\epsfxsize=6.2in
\epsfysize=5.4in
\epsffile{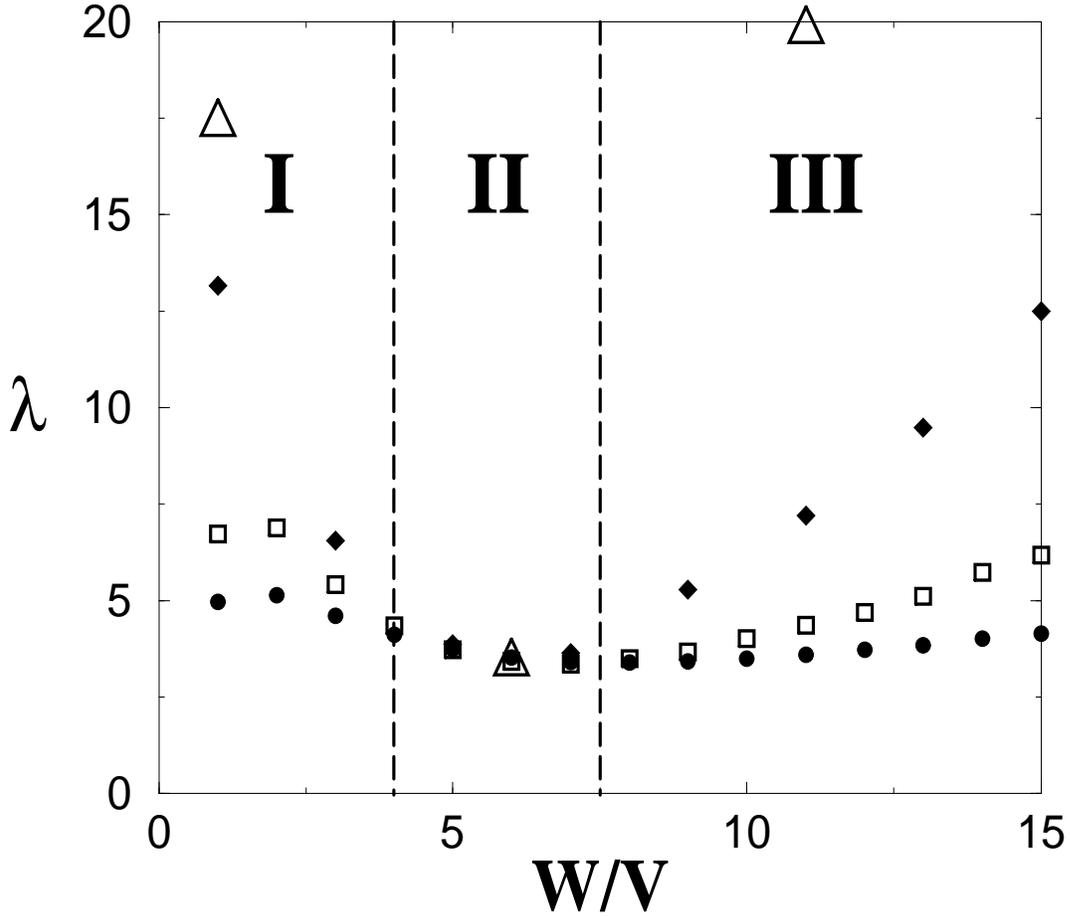}
\caption{Magnetization parameter $\lambda$ vs. disorder strength $W/V$,
for a Hubbard interaction
and a two-dimensional 10$ \times $10 (full circles),
20$ \times $20 (empty squares), 50$ \times $50 (full diamonds) and
80$ \times $80 (empty triangles) Anderson lattice. One clearly
differentiates three regimes : (I) At small disorder, $\lambda$ increases
due to a crossover from ballistic to diffusive behaviour (see text).
(II) At intermediate disorder, exchange
and fluctuations compensate each other so that $\lambda$ is size-independent.
(III) At large
disorder one-body states are strongly localized over very few sites,
which kills the off-diagonal fluctuations faster than
the exchange and the latter dominates again.}
\label{fig:lambda}
\end{figure}

\begin{figure}
\epsfxsize=6.2in
\epsfysize=5.4in
\epsffile{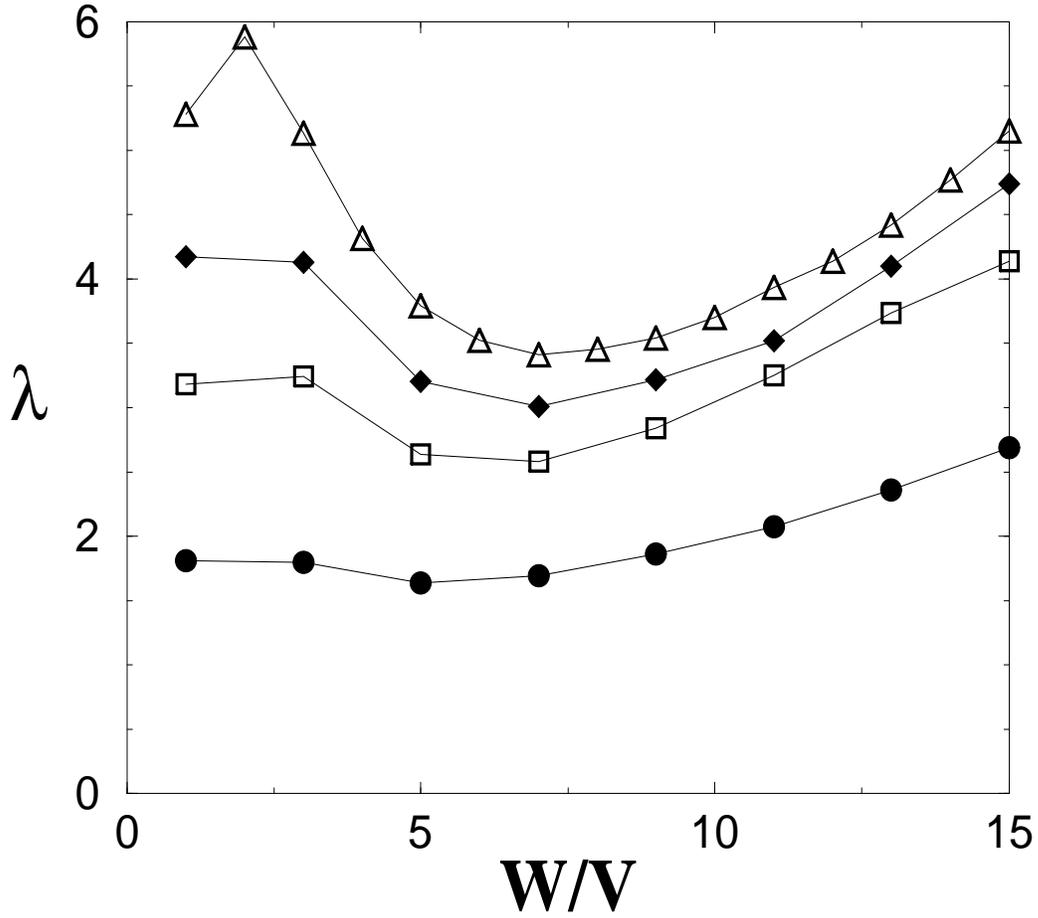}
\caption{Effect of the interaction range on the magnetization
parameter $\lambda$ for a two-dimensional 15$ \times $15
Anderson lattice and ratio $U_0/U_1=1$ (full circles), 4 (empty squares),
9 (full diamonds) and $\infty$ (Hubbard interaction - empty triangles).
As expected, an increase in the interaction range leads to a stronger
increase of the fluctuations than of the exchange, resulting in a lowering
of $\lambda$.}
\label{fig:screening}
\end{figure}

\end{document}